\documentclass[12pt]{iopart}
\usepackage{graphicx}
\usepackage{gensymb}

\begin{document}

\title[LiCrTe\textsubscript{2} single crystals]{Synthesis and Anisotropic Magnetic Properties of LiCrTe\textsubscript{2} Single Crystals with a Triangular-Lattice Antiferromagnetic Structure}

\author{Catherine Witteveen,\textit{$^{1,2}$}  Elisabetta Nocerino,\textit{$^{3}$} Sara A. L\'{o}pez-Paz,\textit{$^{1}$} Harald O. Jeschke,\textit{$^{4}$} Vladimir Y. Pomjakushin,\textit{$^{3}$} Martin M\aa nsson,\textit{$^{5}$} Fabian O. von Rohr\textit{$^{1}$}$^{\ast}$}

\address{$^1$ Department of Quantum Matter Physics, University of Geneva, Quai Ernest-Ansermet 24, 1211 Geneva, Switzerland }
\address{$^2$ Department of Physics, University of Zürich, Winterthurerstr. 190, 8057 Zürich, Switzerland}
\address{$^3$ ~Laboratory for Neutron Scattering and Imaging, Paul Scherrer Institute, CH-5232 Villigen PSI, Switzerland}
\address{$^4$ ~Research Institute for Interdisciplinary Science, Okayama University, Okayama 700-8530, Japan }
\address{$^5$ Department of Applied Physics, KTH Royal Institute of Technology, Roslagstullsbacken 21, SE-106 91 Stockholm, Sweden}
\ead{Fabian.vonRohr@unige.ch}
\vspace{10pt}
\begin{indented}
\item[]March 2023
\end{indented}

\begin{abstract}
We report on the synthesis of LiCrTe\textsubscript{2} single crystals and on their anisotropic magnetic properties. We have obtained these single crystals by employing a Te/Li-flux synthesis method. We find LiCrTe\textsubscript{2} to crystallize in a  TlCdS\textsubscript{2} -type structure with cell parameters of $a$ = 3.9512(5) {\AA} and $c$ = 6.6196(7) {\AA} at $T$ = 175 K. The content of lithium in these crystals was determined to be near stoichiometric by means of neutron diffraction. We find a pronounced magnetic transition at $T^{\rm ab}_{\rm N}$ = 144 K and $T^{\rm c}_{\rm N}$ = 148 K, respectively. These transition temperatures are substantially higher than earlier reports on polycrystalline samples. We have performed neutron powder diffraction measurements that reveal that the long-range low-temperature magnetic structure of single crystalline LiCrTe\textsubscript{2} is an A-type antiferromagnetic (AFM) structure. Our DFT calculations are in good agreement with these experimental observations. We find the system to be easy axis with moments oriented along the $c$-direction experimentally as well as in our calculations. Thereby, the magnetic Hamiltonian can be written as
$H = H_{\rm Heisenberg} + \sum_i K_c (S_i^z)^2$ with $K_c=-0.34K$ (where $|S^z|=\frac{3}{2}$). We find LiCrTe\textsubscript{2} to be highly anisotropic, with a pronounced metamagnetic transition for $H \perp ab$ with a critical field of $\mu H_{MM}$(5 K) $\approx$ 2.5 T. Using detailed orientation-dependent magnetization measurements, we have determined the magnetic phase diagram of this material. Our findings suggest that LiCrTe\textsubscript{2} is a promising material for exploring the interplay between crystal structure and magnetism, and could have potential applications in spin-based 2D devices.
\end{abstract}

%
%
%
%
%

\section{Introduction}
Materials containing a triangular lattice with magnetic interactions have been identified to be a promising route to obtain exotic quantum matter.\cite{Wannier1950, Schmidt2017, Bastien2020} This type of lattice allows for the realization of magnetically frustrated ground states, charge-ordering, and has even been found to host unconventional superconductivity, e.g., in water-intercalated Na\textsubscript{x}CoO\textsubscript{2}.\cite{palmer1984models,mcqueen2008successive,terada2012spiral,toth2016electromagnon,wang2003spin,takada2003superconductivity,schaak2003superconductivity} Many of the prominently discussed triangular lattice compounds crystallize in the delafossite structure with the formula A\textsuperscript{1+}B\textsuperscript{3+}O\textsubscript{2}\textsuperscript{2-}. The delafossite ACrO\textsubscript{2} compounds with Cr(III) in an S=3/2 spin configuration have been extensively studied, resulting in exceptional properties such as e.g., metallic antiferromagnetism in PdCrO\textsubscript{2} or the simultaneous magnetic and ferroelectric orders in AgCrO\textsubscript{2} and CuCrO\textsubscript{2}.\cite{PhysRevLett.105.137201,Takatsu2009, Lopes2011}

Furthermore, the closely related ABX\textsubscript{2}-type compounds with X = S, Se, and Te are currently the object of wide interest. \cite{rasch2009magnetoelastic,carlsson2011suppression,Kobayashi2016,sugiyama2018deviation} These compounds can either be understood as less ionic versions of the delafossites, or they may be regarded as intercalated van-der-Waals transition metal dichalcogenide (TMD) materials. 

Chromium-based van-der-Waals materials have been of intense interest recently, especially the trihalides CrX\textsubscript{3} with X = Cl, I or Br but also the mixed anion compound CrSBr.\cite{huang2017layer,bedoya2021intrinsic,wu2022quasi,Lopez-Paz2022} Accessibility of magnetic TMDs with Cr(IV) cations has been realized by soft-chemical synthesis methodologies, i.e., via indirect oxidation of polycrystalline KCrS\textsubscript{2}, KCrSe\textsubscript{2}, and KCrTe\textsubscript{2}.\cite{Ronneteg2005, Kobayashi2019,song2019soft}. The resulting CrX\textsubscript{2} phases represent promising candidates for magnetic chromium based van-der-Waals materials that have gained much interest for the future realization of quantum technologies like spintronics.\cite{gibertini2019magnetic}

In this respect, the family of ACrX\textsubscript{2} materials with X = S, Se or Te are at the interface between the delafossite-type structures and TMDs. Their properties are insufficiently explored so far, likely due to the unavailability of high-quality single crystals for some of these compounds.\cite{Kobayashi2016, Baenitz2021, Kobayashi2019} Nonetheless, these materials have been observed to display a rich range of magnetic properties. TlCrTe\textsubscript{2}, NaCrSe\textsubscript{2}, and KCrS\textsubscript{2} have been reported to  display long-range AFM order, consisting of ferromagnetic (FM) layers stacked antiferromagnetically\cite{Ronneteg2005, Engelsman1973, Bongers1968, VanLaar1973}. AgCrSe\textsubscript{2} or NaCrS\textsubscript{2} have been found to display an AFM helical order, and LiCrSe\textsubscript{2} shows a complex up-up-down-down magnetic structure with itinerant suppression of the Cr moment.\cite{Engelsman1973, Bongers1968, Nocerino2022a}

In this work, we report on the synthesis of high-quality single crystals of LiCrTe\textsubscript{2}, which we have characterized with a series of magnetization and neutron diffraction measurements. We show orientation dependent magnetism as a result of interlayer coupling and established a magnetic phase diagram for the observed FM spin-rearrangements. We believe that the availability of single crystals of LiCrTe\textsubscript{2} is an important step in understanding and experimentally showing the impact of alkali metal intercalation on the electronic and magnetic properties of layered TMDs.

\section{Experimental Section}
Single crystals were prepared using a Te/Li flux. Lithium (granulates, Sigma Aldrich, 99\%), chromium (powder, Alfa Aesar, 99.99\%), and tellurium (pieces, Alfa Aesar, 99.999\%) were used as received and mixed in the ratio 3.3:1:8. The elements were placed in an alumina Canfield crucible set consisting of a bottom and top crucible separated by a frit-disc, which was subsequently sealed in a quartz ampule under a 1/3 atm argon atmosphere after having been purged three times.\cite{Canfield2016} The ampules were then heated in a muffle furnace (heating rate 30 \degree C/h) to 1000\degree C and slowly cooled to 550 \degree C over 4 days (96 hours). The samples were removed from the furnace and immediately centrifuged, in order to separate the crystals from the flux.   

Powder X-ray diffraction (PXRD) patterns were collected on thin-plate-like crystals laying flat (parallel to the $ab$-plane) using a Rigaku SmartLab in reflection mode with Cu K\textsubscript{$\alpha$} radiation. To avoid fast decomposition, the crystals were covered using paraffin oil and measured in an airtight sample holder.  

The temperature-dependent and field-dependent magnetization was measured in a Quantum Design magnetic properties measurement system (MPMS 3)  equipped with a 7 T magnet and with the vibrating sample magnetometer (VSM). The samples were measured in a gelatin capsule prepared in an argon glovebox. Depending on the specific measurement, the crystal directions were oriented accordingly.

Neutron diffraction patterns were obtained on the High-Resolution Powder Diffractometer (HRPT) at the Swiss Spallation Neutron Source (SINQ) from the Paul Scherrer Institute (PSI) in Villigen, Switzerland. Single crystals were finely powderized in an argon filled glovebox using an agate mortar and a file. The powder was then transferred to a vanadium sample container of 8 mm diameter and 50 mm length and sealed with an indium ring under helium atmosphere. Data were collected at $T$ = 1.8 K and 175 K with a neutron wavelength of 1.886 {\AA} for magnetic structure refinements and structural refinements (175 K) respectively and analyzed by the Rietveld method using the Fullprof Suite package. The magnetic symmetry analysis was done using ISODISTORT from the ISOTROPY software and BasIreps from the Fullprof suite.\cite{Stokes, Campbell2006} The peak shape was modeled using a Thompson–Cox–Hastings pseudo-Voigt function with axial divergence asymmetry (as implemented in Fullprof with $Npr=7$). The preferential orientation characteristic of layered materials was observed in the neutron diffraction patterns and was refined using the modified March’s function (as implemented in Fullprof with $Nor=1$).

The Hamiltonian of LiCrTe\textsubscript{2} (and NaCrTe\textsubscript{2}) was determined by density functional theory-based energy mapping.\cite{Jeschke2015,Jeschke2019} We use the all electron  full potential local orbital (FPLO) code~\cite{Koepernik1999} for all density functional theory calculations, in combination with the
generalized gradient approximation (GGA) exchange and correlation
functional.\cite{Perdew1996} We base our calculations both on the neutron diffraction crystal structures determined in this work and on the structures reported in reference [\cite{Kobayashi2016}]. We account for strong correlations on the Cr $3d$ orbitals by applying a GGA+U exchange correlation functional~\cite{Liechtenstein1995} for several different values of $U$ and $J_H=0.72$\,eV fixed following reference \cite{Mizokawa1996}.
We are fitting to the Heisenberg Hamiltonian in the form
$H=\sum_{i<j} J_{ij} {\bf S}_i\cdot {\bf S}_j$. Total moments are
exact multiples of 3\,$\mu_{\rm B}$ as all chromium spins are exactly $S=3/2$. The Curie-Weiss temperature estimates
are obtained from
\begin{equation*}
 \theta_{\rm CW} = -\frac{2}{3} S (S + 1) (3J_1+ J_2 + 3J_3 + 6J_4 + 3J_5 + 6J_6 +6J_7)
\end{equation*}
where $S=3/2$. For the determination of the magnetocrystalline anisotropy energy, we have accounted for spin-orbit coupling using fully relativistic total energy calculations.

\section{Result and Discussion}
\subsection{Synthesis and Crystal Structure of LiCrTe\textsubscript{2} Single Crystals }

\begin{figure}
\centering
  \includegraphics[width=0.7\linewidth]{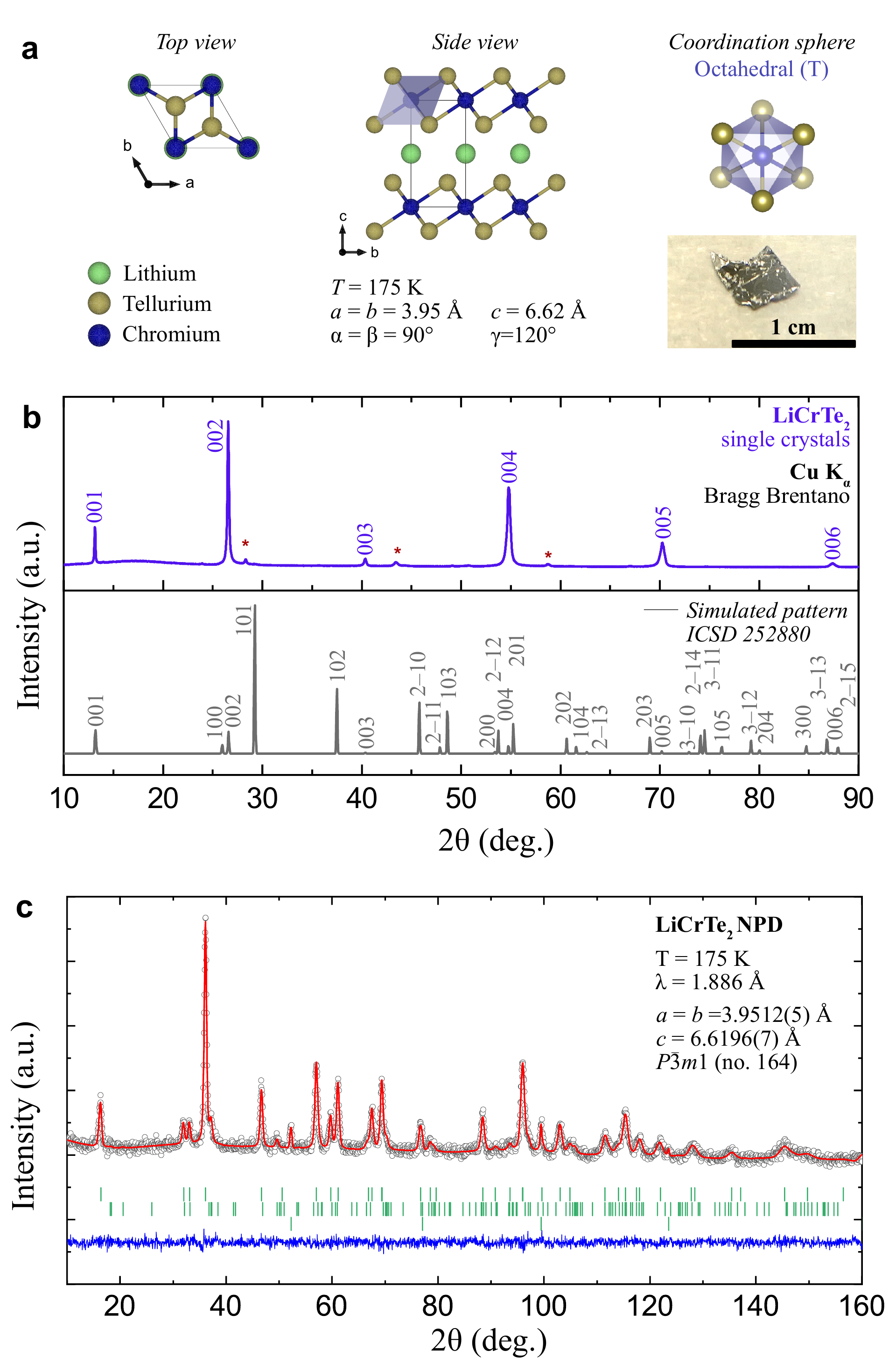}
  \caption{Single crystal growth and structure of LiCrTe\textsubscript{2}. (a) Left top: Schematic representation of the crystal structure along the $c$-axis, showing the octahedral coordination and the number of  CrTe\textsubscript{2}  layers (1) per unit cell. In green lithium (Li), in yellow tellurium (Te) and in blue chromium (Cr). Middle: Representation along the $a$-axis. Right: octahedral coordination of  CrTe\textsubscript{6}  and a photograph of a typical as-grown crystal. (b) Measured PXRD pattern of single crystals using a Bragg-Brentano geometry in blue with Cu \(K_{\alpha}\) radiation, including the respective Miller indices of the $(\textit{00l})$ reflections and simulated PXRD patterns of LiCrTe\textsubscript{2} in gray. Impurities (red stars) could be assigned to  Cr\textsubscript{2}Te\textsubscript{3}  and are also apparent in the Neutron Powder Diffraction (NPD). (c) NPD at $T$ = 175 K, above the AFM transition, especially used for the determination of the respective Li content, with the corresponding Rietveld refinement for a lithium content of $\ge $ 92 \%. The secondary phase is  Cr\textsubscript{2}Te\textsubscript{3}, the ternary phase belongs to vanadium from the sample container.}
  \label{Figure1}
\end{figure}

Using a Te/Li self flux, we have synthesized single crystals of LiCrTe\textsubscript{2}. The crystals are silvery shiny and of metallic luster. Large plate-like crystals up to a dimension of 8 mm x 8 mm x 1 mm were extracted. The crystals are found to be highly air sensitive, as they decompose within seconds in air.
\\
LiCrTe\textsubscript{2} crystallizes in the  TlCdS2\textsubscript{2} structure type and belongs to the space group no. 164 ($P\bar{3}m1$), like the isomorphic NaCrTe\textsubscript{2} or KCrTe\textsubscript{2}.\cite{Kobayashi2016, Sun2020} This compound can be understood as a Li intercalated version of the thermodynamically metastable 1T-CrTe\textsubscript{2} van der Waals compound. In figure \ref{Figure1}(a), a schematic representation of the crystal structure is depicted with the view along the $a$- and $c$- directions with the unit cell drawn in black. Furthermore, the octahedral coordination of CrTe\textsubscript{6} polyhedra is depicted and a photograph of a typical as-grown single crystal of LiCrTe\textsubscript{2} is shown. 
In figure \ref{Figure1}(b), we show the measured PXRD of the synthesized crystals (purple) using a Bragg-Brentano geometry. Due to the orientation of the crystal with the $c$-axis perpendicular to the sample holder, only the reflections of the (00l) planes are observed. The simulated PXRD pattern of LiCrTe\textsubscript{2} is presented below for comparison. We observe impurities (red stars) that are in good agreement with the (00l) reflections of Cr\textsubscript{2}Te\textsubscript{3}. The impurity is also apparent in the neutron powder diffraction (NPD).
In figure \ref{Figure1}(c), NPD data of powderized single crystals of LiCrTe\textsubscript{2} is shown, along with its corresponding Rietveld refinement. The crystal structure previously proposed by \textit{Kobayashi et al.} -- obtained from polycrystalline samples using PXRD measurements -- could be confirmed from our Rietveld refinements of the neutron diffraction data collected at $T$= 175 K.\cite{Kobayashi2016} 

 We find LiCrTe\textsubscript{2} to be well-described in the space group 164 ($P\bar{3}m1$) with the unit cell parameters $a$ = 3.9512(5) {\AA}  and $c$ = 6.6196(7) {\AA} at $T$ = 175 K. We observe a smaller value for the $a$-axis on the single crystals $T$ = 175 K compared to $a$ = 3.96 {\AA} from Kobayashi et al. at 295 K.\cite{Kobayashi2016}. Impurity peaks observed in the neutron diffraction data could be assigned to   Cr\textsubscript{2}Te\textsubscript{3}  as well as vanadium from the sample container. The details of the NPD Rietveld refinement are shown in table \ref{table1}.

The lithium content was determined by freeing the occupation parameter of the respective Wyckoff site. An occupancy of $\ge$ 92 \% was found to give the best result for this refinement. Manually lowering the content of lithium in the refinement would drastically decrease the intensity of the (001), (100), (002), and (101) reflections around 2$\theta$ = 16-40\degree, strongly hinting at a near nominal content of lithium in the single crystals. \\

\subsection{Long-Range Magnetic Order in LiCrTe\textsubscript{2} Single Crystals }

\begin{figure}
  \centering
  \includegraphics[width=0.7\linewidth]{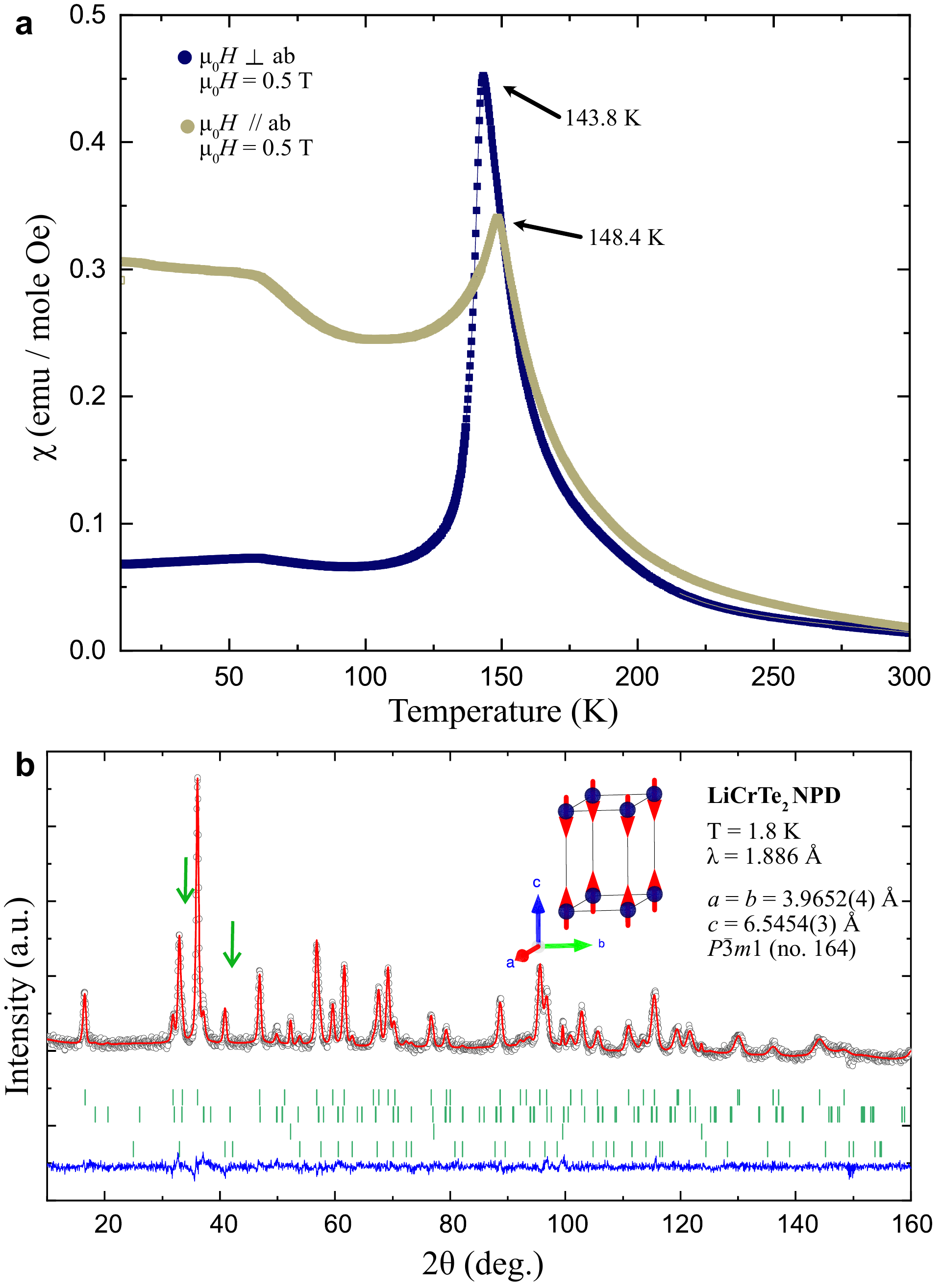}
  \caption{AFM order in LiCrTe\textsubscript{2} single crystals: (a) Magnetic susceptibility of LiCrTe\textsubscript{2} single crystals in-plane (beige) and out-of-plane (blue) measured in an external field of $\mu_0 H$ = 0.5 T in a temperature range between $T$ = 10 K and 300 K. (b) Neutron powder diffraction (NPD) data of powderized LiCrTe\textsubscript{2} single crystals at $T$ = 1.8 K, well below the AFM transition. The green arrows highlight the magnetic reflections. Inset: Graphical representation of the A-type AFM structure in single crystalline LiCrTe\textsubscript{2}, the red arrows in the inset represent the ordered Cr moments.}
  \label{Figure2}
\end{figure}

Previously, it has been reported that polycrystalline LiCrTe\textsubscript{2} undergoes a phase transition to a long-range AFM ordered state at a N\'eel temperature of $T_{\rm N}$ = 71 K.\cite{Kobayashi2016} Here, we found that single crystals of LiCrTe\textsubscript{2} also undergo a transition to an AFM state but at a much higher temperature of $T_{\rm N}^{ab}$ = 143.8 K for $\mu_0 H \perp ab$ and $T_{\rm N}^{c}$ = 148.4 K for $\mu_0 H \perp c$ respectively, see figure \ref{Figure2}(a). We find an increase in the AFM state for both orientations culminating at $T \approx$ 60 K, which hints towards an additional low temperature magnetic order, potentially similar to related chromium based layered magnetic materials (see, e.g., references \cite{Lopez-Paz2022, Nocerino2022a}). The elucidation of a possible underlying order (e.g. a spin canted phase) in LiCrTe\textsubscript{2} at low temperature therefore deserves further studies. For example, small angle neutron scattering or resonant X-ray scattering experiments could help accessing a possible short range ordered state.

A tentative fit of the inverse of the susceptibility shows a positive Curie-Weiss temperature ($\approx$ 160 K) indicating a FM direct coupling within the chromium sheets, in contrast to the AFM ordering along the $c$ axis, see SI.

These results from the magnetic susceptibility measurements are in good agreement with an A-type AFM order, as evidenced from our neutron diffraction measurements (see below). 

In order to determine the magnetic structure of LiCrTe\textsubscript{2} single crystals, we have performed low-temperature neutron diffraction experiments, which are depicted in figure \ref{Figure2}(b). The neutron diffraction experiment conducted at $T$ = 175 K (above the AFM transition, see figure \ref{Figure1}(c)) and $T$ = 1.8 K, see figure \ref{Figure2}(b), show a decrease of the $c$ axis but an increase of the $a$ axis on lowering the temperature, indicating a shortening of the Cr-Te, Li-Te and Te-Te distances but an increase in the Cr-Cr distance, see table \ref{table1}. This negative thermal expansion observed is consistent with the recent results of synchrotron data on polycrystalline LiCrTe\textsubscript{2} conducted by Nocerino et al, and has been observed in other intercalated chromium tellurides.\cite{Nocerino2022, Ronneteg2005}

Also a similar $T_{\rm N}$ $\approx$ 125 K from $\mu^+$SR measurements was observed for the powder sample from Nocerino et al.\cite{Nocerino2022}. Considering the stoichiometric Li-content of the powder sample, the lower critical temperature derived from the $\mu^+$SR measurements could reflect the strong dynamic character of the magnetic interactions, or the enhanced $T_{\rm N}$ $\approx$ 148 K in the single crystals might be related to local strain effects in either sample. A direct comparison of these two stoichiometric Li samples remain however qualitative, as different techniques were employed for the determination of the $T_{\rm N}$. The lower $T_{\rm N}$ of \textit{Kobayashi et al.} was extracted at a high external field of $\mu\textit{H}_0$ = 7 T.\cite{Kobayashi2016} Their measurement at the lower $\mu\textit{H}_0$ = 1 T showed no typical AFM cusp. The different observations in those different samples hint towards a versatile system, where the $T_{\rm N}$ is easily tunable by chemical manipulation through strain or Li-content.

The magnetic structure refinement was performed on the $T =$ 1.8 K diffraction pattern to obtain the long-range magnetic structure of LiCrTe\textsubscript{2} single crystal. The emerging magnetic reflections are highlighted by green arrows. The LiCrTe\textsubscript{2} magnetic phase includes a single magnetic Cr atom in its crystallographic site. The scale factor and structural parameters were constrained to be equal to their counterparts in the nuclear LiCrTe\textsubscript{2} phase, in order to obtain a correct value for the Cr magnetic moment of $\mu_{Cr}$ = 3.09 $\pm$ 0.02 $\mu_{B}$. This value is in excellent agreement with what is expected for a Cr(III) ion.

The magnetic propagation vector $q$ for correct indexing of the temperature dependent magnetic Bragg reflections was determined as $q$ = (0 0 1/2) with the software K-Search. The possible irreducible representations (irrep) of the propagation vector group G$_k$ were obtained from the paramagnetic parent space group $\textit{P}\overline{3}\textit{m}1$ and the calculated propagation vector $q$ with the software BasIreps, resulting in irrep mA\textsuperscript{2+} according to the international classification of magnetic superspace group. These results are consistent with the result obtained from ISODISTORT from the ISOTROPY software.\cite{Stokes, Campbell2006} Both K-search and BasIreps are part of the FullProf suite of programs for refinement.\cite{Rodriguez-Carvajal1993}

Among the possible resulting irreps, the magnetic structure refinement with irrep $\Gamma_3$, corresponding to the Shubnikov Group $P_c\overline{3}c1$ (nr. 165.96), was the only one able to well capture the temperature dependent magnetic features of the diffraction pattern. This irrep consists of a single basis vector with no imaginary component BsV: Re (0 0 1). We find the magnetic structure of single crystalline LiCrTe\textsubscript{2} to be A-type antiferromagnetic, with the Cr moments oriented along the $c$-axis. This result is in excellent agreement with earlier findings in polycrystalline LiCrTe\textsubscript{2}, where the same magnetic structure was found.\cite{Nocerino2022} The cell parameters and magnetic moment resulting from the refinement are summarized in table \ref{table1} together with the Bragg $R$ agreement factors for the nuclear and magnetic LiCrTe\textsubscript{2} phases.

\begin{table}
\caption{Comparison of results from neutron powder diffraction on single crystalline LiCrTe\textsubscript{2} at two different temperatures. Spacegroup no. 164 ($P\bar{3}m1$)}
  \label{table1}
  \begin{tabular*}{\textwidth}{@{\extracolsep{\fill}}  c c c c}
    \hline
                &              & \textbf{175 K}       & \textbf{1.8 K }      \\
    \hline
    
          & \textit{a} (\AA)     & 3.9512(5)  & 3.9652(4)                   \\
         & \textit{c} (\AA)     & 6.6196(7)  & 6.5454(3)                   \\
         & \textit{V} (\AA \textsuperscript{3})    &  89.502(2) & 88.125(1)  \\
         \\
                & Cr-Te (\AA)    & 2.743(3)  & 2.737(2)     \\
                & Li-Te (\AA)    & 2.897(3)  & 2.895(2)    \\
                & Te-Te  (\AA)   & 3.807(7)  & 3.774(5)    \\
                & Cr-Cr (\AA)   & 3.9512(5)  & 3.9652(4)     \\
    \hline
           Atom          & Site; Wyckoff pos         &           &       \\
    \hline
           Li            & (0, 0, $\frac{1}{2}$); 1\textit{b}  &           &       \\
                        & occ & 0.078(3)           &   0.078(3)    \\
                         & \textit{B}\textsubscript{\textit{iso}} (\AA \textsuperscript{3}) & 2.77(50) & 2.13(22)       \\
    \hline
           Cr            & (0, 0, 0); 1\textit{a}  &           &       \\
                        & occ &   0.0833(0)        &   0.0833(0)     \\
                         & \textit{B}\textsubscript{\textit{iso}} (\AA \textsuperscript{3}) & 1.14(20) & 0.76(1)       \\   
    \hline
           Te            & ($\frac{1}{3}$, $\frac{2}{3}$, \textit{z}); 2\textit{d}  &           &       \\
                         & \textit{z}                    & 0.2316(7)        & 0.2302(5) \\
                & occ  &     0.167(3)      &   0.167(3)    \\
                         & \textit{B}\textsubscript{\textit{iso}} (\AA \textsuperscript{3}) & 1.10(7) & 0.66(4)     \\
    \hline
                         & $\mu_{Cr}$           & -       & 3.09(2)\\
                         & $R_{B(cryst)}$ (\%)  & 2.69       & 2.38        \\
                         & $R_{B(mag)}$ (\%)    & -         & 12.4        \\
                         & $\chi\textsuperscript{2}$   &  1.16       & 1.5  \\
                         
    \hline
  \end{tabular*}
\end{table}

\subsection{Magnetic and Electronic Structure Calculations} 

We have investigated the electronic structure and magnetic interactions of LiCrTe\textsubscript{2} using all electron density functional theory calculations. In the environment of the  CrTe\textsubscript{6}  octahedra, Cr $3d$ states split into half-filled $t_{2g}$ states and unoccupied $e_g$ states. Spin-polarized calculations show that the Cr$^{3+}$ ions have $S=\frac{3}{2}$ moments, and at $U=1$\,eV the system is insulating with a gap of $E_{\rm g}=0.48$\,eV. We use the energy mapping technique that has yielded very good results in other chromium-based magnets~\cite{Ghosh2019} to extract the Heisenberg Hamiltonian parameters.
 
 \begin{figure}[h]
\centering
  \includegraphics[width=1\linewidth]{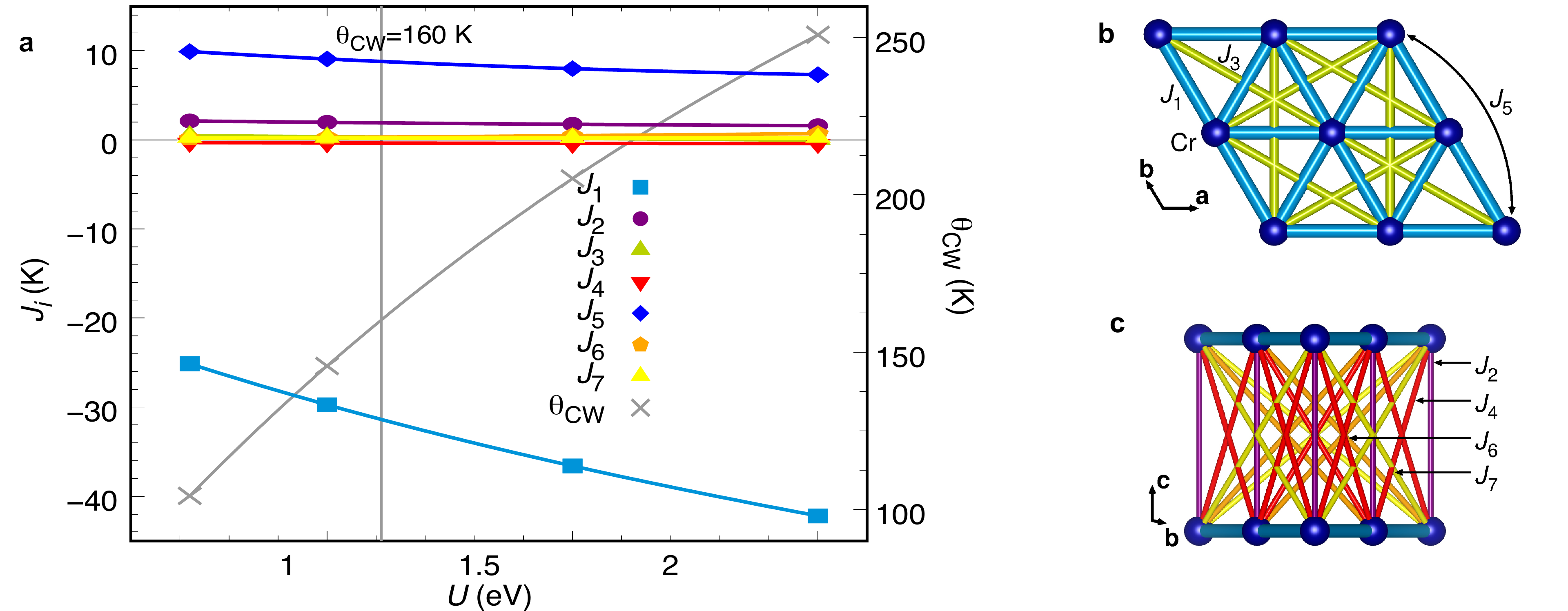}
  \caption{Hamiltonian parameters of LiCrTe\textsubscript{2}. (a) Exchange couplings of LiCrTe\textsubscript{2} determined by DFT energy mapping as function of onsite interaction strength $U$ for the $T=175$\,K structure. The vertical line indicates the $U$ value where the exchange couplings match the experimental Curie-Weiss temperature. The resulting exchange couplings are listed in table~\ref{table2}. (b) In-plane exchange paths of LiCrTe\textsubscript{2}. (c) Interlayer exchange paths.}
  \label{Figure3}
\end{figure}
 
Figure \ref{Figure3}(a) shows the result of these calculations for the $T=175$\,K structure extracted from neutron data. Exchange couplings evolve smoothly with the on-site Coulomb repulsion $U$, and the Hamiltonian reproduces the experimental Curie-Weiss temperature at $U=1.1$\,eV (vertical line), in good agreement with other Cr$^{3+}$ compounds~\cite{Ghosh2019}. The dominant coupling in the triangular lattice material LiCrTe\textsubscript{2} is a FM nearest neighbor coupling $J_1$ (see figure \ref{Figure3}(b). The second-largest coupling, $J_5$ which corresponds to twice the $J_1$ distance, is AFM. While it has no effect on the ferromagnetic ordering of the Cr planes, it will affect the spin wave dispersion. Table~\ref{table2} shows the in-plane couplings for structures of LiCrTe\textsubscript{2} determined at three different temperatures. There is very little change of the Hamiltonian between $T=1$\,K and room temperature. For comparison, the Hamiltonian for  NaCrTe\textsubscript{2} , calculated at $U=1$\,eV, is also shown. Among the many interlayer couplings of LiCrTe\textsubscript{2} (see figure \ref{Figure3} c), only $J_2$ has an appreciable value and is AFM. In order to be sure about the sign of the cumulative effect of the interlayer couplings, we have determined the effective interlayer couplings as $J_\perp=2.8(1)$\,K and $J_{\perp2}=-0.13(1)$\,K ($J_{\perp2}$ connects every second plane along $c$). This confirms that the DFT prediction for LiCrTe\textsubscript{2} is A-type AFM order.

We have also calculated the magnetocrystalline anisotropy energy by determining the energy of the FM configuration for different orientations of the magnetic moments with spin-orbit coupling taken into account. We find the system to be easy axis with moments oriented along the $c$ direction. The Hamiltonian can be written as
$H = H_{\rm Heisenberg} + \sum_i K_c (S_i^z)^2$ with $K_c=-0.34K$ (where $|S^z|=\frac{3}{2}$).
In the $ab$ plane, all orientations have the same energy. All calculations are in excellent agreement with the observed experimental results.

\begin{table}

  \caption{In-plane exchange couplings of LiCrTe\textsubscript{2} and  NaCrTe\textsubscript{2} , calculated within GGA+U and $6\times 6\times 6$ $k$ points. The experimental value $T_{\rm CW}=160$\,K of the Curie Weiss temperature is matched by the couplings for LiCrTe\textsubscript{2}. Statistical errors are indicated.}
 
  \label{table2}

  \begin{tabular*}{\textwidth}{@{\extracolsep{\fill}}  c c c c}

    \hline
Material & $J_1$ (K) &    $J_3$  (K)&    $J_5$ (K)  \\
    \hline
LiCrTe\textsubscript{2}, $T=1$  K & -32.4(1) & 0.7(1) & 9.5(1)\\
LiCrTe\textsubscript{2}, $T=175$  K &-31.4(1)&0.3(1)&8.8(1)\\
LiCrTe\textsubscript{2}, $T=300$ K\cite{Kobayashi2016} & -30.9(1) & 0.0(1) & 8.8(1)\\
NaCrTe\textsubscript{2} , $T=300$ K\cite{Kobayashi2016} & -39.9(1) & 1.0(1) & 9.2(1)\\
\end{tabular*}

\end{table}

\subsection{Magnetic Anisotropy and Phase Diagram}

\begin{figure*}
\centering
  \includegraphics[width=1\linewidth]{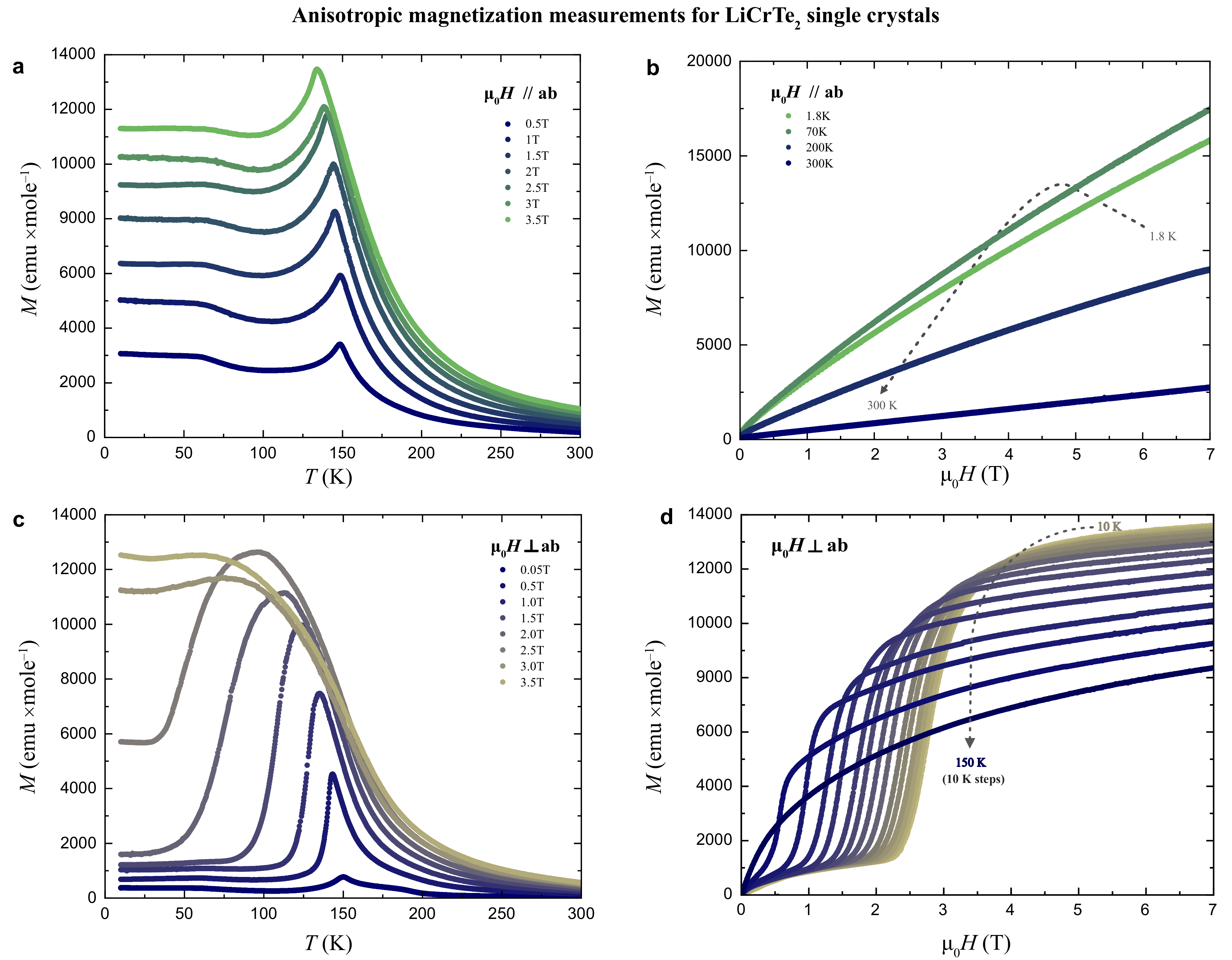}
  \caption{Temperature dependent and field dependent magnetization for a single crystal with the $ab$ plane parallel (a, b) and perpendicular (c, d) to the external magnetic field. In (c) we observe a transition to FM spin orientation with the crystal undergoing a metamagnetic transition with an increasing broadening of the transition with the applied field perpendicular to the crystal $ab$ plane. This behaviour is not observed when the ab plane lies parallel to the applied field as in (a) in the field dependent magnetization.}
  \label{Figure4}
\end{figure*}

The magnetic anisotropic properties of LiCrTe\textsubscript{2} were investigated by means of field- and temperature-dependent magnetic property measurements. In figure \ref{Figure4}(a) and (c) we show the temperature-dependent magnetization measured with the magnetic field parallel and perpendicular to the $ab$-plane, respectively. The magnetization is shown in a temperature range between \textit{T}= 10 K and 300 K and in magnetic fields between $\mu_0 H$ = 0.5 to 3.5 T. In figure \ref{Figure4}(b) and (d), we show the corresponding field-dependent magnetization measurements (b) parallel to the $ab$-plane and (d) perpendicular to the $ab$-plane, respectively. These measurements were performed at temperatures between $T$ = 1.8 K and 300 K in magnetic fields between $\mu_0 H$ = 0 to 7 T.

We find that when the $ab$-plane is perpendicular to the applied magnetic field, a pronounced metamagnetic transition occurs. This field-influenced transition can be interpreted as a transition from the long-range magnetically ordered AFM state to a long-range magnetically ordered state with all the magnetic moments oriented in the same direction, equivalent to a long-range FM order. The metamagnetic transition is strongly pronounced both in the temperature-dependent, as well as the field-dependent measurements. The transition has fully occurred above a field of $\mu_0 H$ = 3.5 T. No field-induced magnetic transition is found for external fields that are applied in parallel to the $ab$-plane for fields up to $\mu_0 H$ = 7 T. The critical temperature has only a weak field-dependence for both orientations.

\begin{figure}
\centering
  \includegraphics[width=0.7\linewidth]{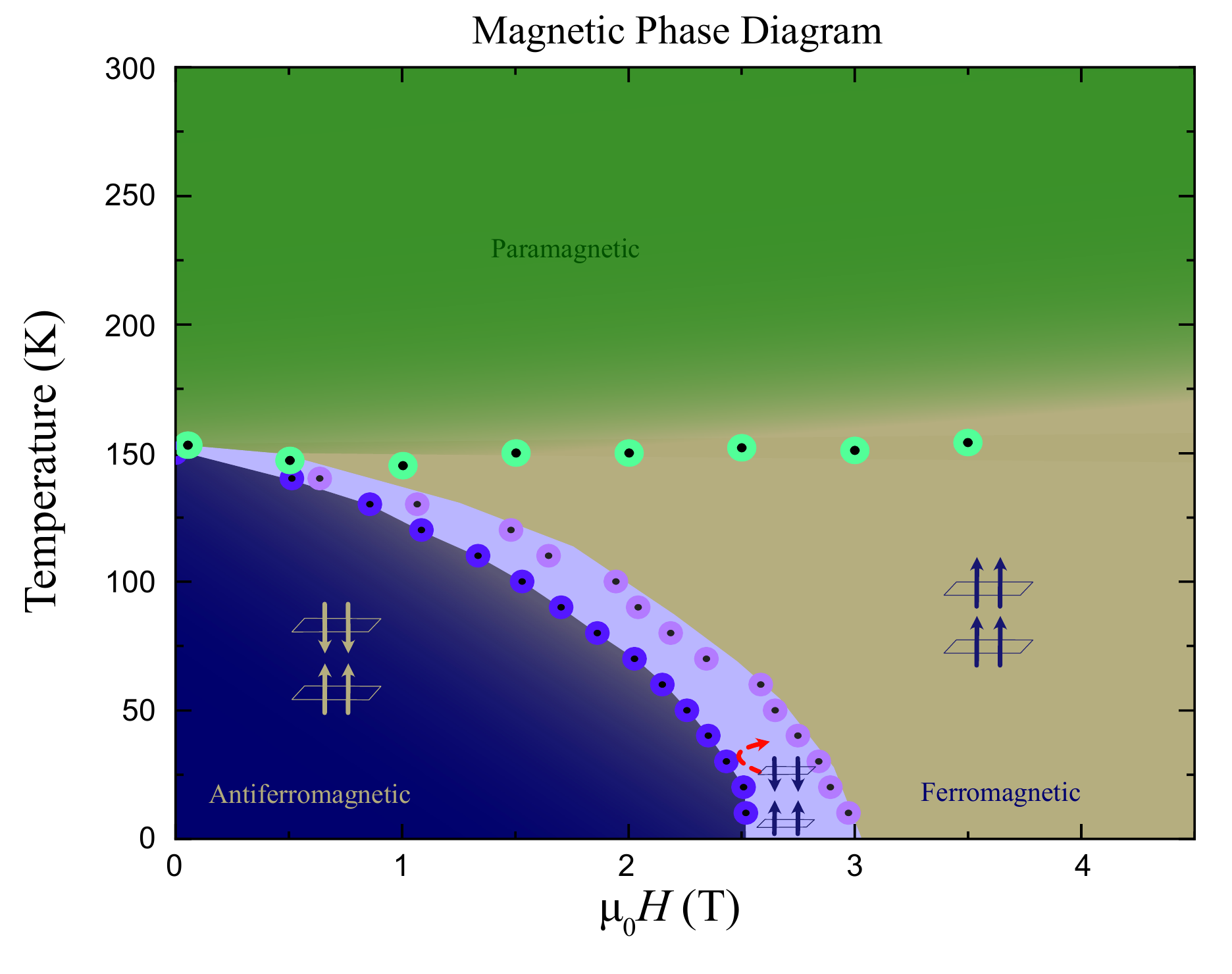}
  \caption{Magnetic phase diagram of LiCrTe\textsubscript{2} in external magnetic fields perpendicular to the $ab$-plane in a field range between $\mu_0H =$ 0 and 4.5 T and in a temperature range between $T$ = 0 and 300 K. The MM transition was determined from the 1$^{\rm st}$ derivative of the magnetization in figure \ref{Figure4}(d) (dark purple and light purple  circles) and the FM transition from the first derivative of the magnetization in figure \ref{Figure4}(c) (light green circles)}
  \label{Figure5}
\end{figure}

Based on the results shown in figure \ref{Figure4}(c) and (d), we established the magnetic phase diagram for the LiCrTe\textsubscript{2} single crystals perpendicular to the applied magnetic field. Four different magnetic phases are identified: a paramagnetic high-temperature phase above the $T_{\rm N}$ (PM, green), an AFM ordered phase (AFM, blue), the metamagnetic phase transition (MM, purple) where the spins flip from AFM to FM order, and a FM ordered phase (FM, brown). We show the magnetic phase diagram in figure~\ref{Figure5}. We have used the first derivative of the magnetization data in figure \ref{Figure4}(c) for the FM transition $T_{\rm C}$ (light green  circles) and the deviations from linearity observed in the first derivative of the magnetization in figure \ref{Figure4}(d) as a measure for $H_{\rm MM}$ (dark purple and light purple  circles). The detailed procedure is illustrated in the SI.

The observed metamagnetic transition in this material is an addition to the already mentioned induced functionality change upon lithium adsorption. Single crystals of LiCrTe\textsubscript{2} stand as an interesting system, offering many approaches to tune its property by intercalation/deintercaltation and application of external fields.


\section{Summary and Conclusions}
In summary, we report on the synthesis and the magnetic properties of LiCrTe\textsubscript{2} single crystals. We have performed the synthesis of LiCrTe\textsubscript{2} single crystals using a Te/Li self flux method. Using neutron powder diffraction and by SQUID magnetometry measurements, we have identified the magnetic properties of this material. We found an AFM transition temperature of $T_{\rm N} \approx$ 148 K by magnetic susceptibility measurements. This transition temperature is substantially higher than the transition temperature reported earlier for polycrystalline samples of LiCrTe\textsubscript{2}.\cite{Kobayashi2016} From our neutron diffraction measurements, we were able to determine the magnetic structure of single crystalline LiCrTe\textsubscript{2} as a collinear magnetic structure with a FM configuration within the  CrTe\textsubscript{2}  layers and the spins lying along the $c$-axis. Thereby, the  CrTe\textsubscript{2}  layers are ordered ferromagnetically and the layers are coupled antiferromagnetically along the $c$-axis. Our experimental results can be well described by DFT calculations with a Hamiltonian that can be written as $H = H_{\rm Heisenberg} + \sum_i K_c (S_i^z)^2$ with $K_c=-0.34K$ (where $|S^z|=\frac{3}{2}$). Both experimental evidence as well as the DFT calculations confirm the A-type AFM order in LiCrTe\textsubscript{2}, where the magnetic moments align ferromagnetically in the \textit{ab}-plane and the planes order antiferromagnetically along the \textit{c}-axis. Furthermore, we find a field-induced magnetic transition for magnetic fields applied perpendicular to the \textit{ab}-plane from the long-range AFM to an FM state in LiCrTe\textsubscript{2}. Using detailed orientation-dependent magnetization measurements, we were able to establish a magnetic phase diagram for this system for fields applied perpendicular to the \textit{ab}-plane.

\section*{Conflicts of interest}
There are no conflicts to declare.

\section*{Acknowledgements}
The authors thank Robin Lefèvre and Johan Chang for insightful discussions. This  work  was  supported by the Swiss National Science Foundation under Grant No. PCEFP2\_194183. E.N. is supported by the Swedish Foundation for Strategic Research (SSF) within the Swedish national graduate school in neutron scattering (SwedNess), as well as the Swedish Research Council VR (Dnr. 2021-06157 and Dnr. 2017-05078)

\section*{References}
\bibliography{XCrTe2.bib} 

\providecommand{\newblock}{}
\begin{thebibliography}{10}
\expandafter\ifx\csname url\endcsname\relax
  \def\url#1{{\tt #1}}\fi
\expandafter\ifx\csname urlprefix\endcsname\relax\def\urlprefix{URL }\fi
\providecommand{\eprint}[2][]{\url{#2}}

\bibitem{Wannier1950}
Wannier G~H 1950 {\em Phys. Rev.\/} {\bf 79}(2) 357--364

\bibitem{Schmidt2017}
Schmidt B and Thalmeier P 2017 {\em Phys. Rep.\/} {\bf 703} 1--59
  (\textit{Preprint} \eprint{1710.04399})

\bibitem{Bastien2020}
Bastien G, Rubrecht B, H{\"{a}}u{\ss}ler E, Schlender P, Zangeneh Z, Avdoshenko
  S, Sarkar R, Alfonsov A, Luther S, Onykiienko Y~A, Walker H~C, K{\"{u}}hne H,
  Grinenko V, Guguchia Z, Kataev V, Klauss H~H, Hozoi L, van~den Brink J,
  Inosov D~S, B{\"{u}}chner B, Wolter-Giraud A and Doert T 2020 {\em SciPost
  Phys.\/} {\bf 9} 1--20 (\textit{Preprint} \eprint{2005.06300})

\bibitem{palmer1984models}
Palmer R~G, Stein D~L, Abrahams E and Anderson P~W 1984 {\em Phys. Rev.
  Lett.\/} {\bf 53} 958

\bibitem{mcqueen2008successive}
McQueen T, Stephens P, Huang Q, Klimczuk T, Ronning F and Cava R~J 2008 {\em
  Phys. Rev. Lett.\/} {\bf 101} 166402

\bibitem{terada2012spiral}
Terada N, Khalyavin D~D, Manuel P, Tsujimoto Y, Knight K, Radaelli P~G, Suzuki
  H~S and Kitazawa H 2012 {\em Phys. Rev. Lett.\/} {\bf 109} 097203

\bibitem{toth2016electromagnon}
T{\'o}th S, Wehinger B, Rolfs K, Birol T, Stuhr U, Takatsu H, Kimura K, Kimura
  T, R{\o}nnow H~M and R{\"u}egg C 2016 {\em Nature Comm.\/} {\bf 7} 1--7

\bibitem{wang2003spin}
Wang Y, Rogado N~S, Cava R~J and Ong N~P 2003 {\em Nature\/} {\bf 423} 425--428

\bibitem{takada2003superconductivity}
Takada K, Sakurai H, Takayama-Muromachi E, Izumi F, Dilanian R~A and Sasaki T
  2003 {\em Nature\/} {\bf 422} 53--55

\bibitem{schaak2003superconductivity}
Schaak R~E, Klimczuk T, Foo M~L and Cava R~J 2003 {\em Nature\/} {\bf 424}
  527--529

\bibitem{PhysRevLett.105.137201}
Takatsu H, Yonezawa S, Fujimoto S and Maeno Y 2010 {\em Phys. Rev. Lett.\/}
  {\bf 105}(13) 137201

\bibitem{Takatsu2009}
Takatsu H, Yoshizawa H and Maeno Y 2009 Comparative study of conductive
  delafossites with and without frustrated spins on a triangular lattice,
  $pdmo_2$ (m = cr, co) {\em Journal of Physics: Conference Series\/} vol 145
  (IOP Publishing) p 012046

\bibitem{Lopes2011}
Lopes A~M, Oliveira G~N, Mendon{\c{c}}a T~M, Moreira J~A, Almeida A,
  Ara{\'{u}}jo J~P, Amaral V~S and Correia J~G 2011 {\em Phys. Rev. B\/} {\bf
  84} 1--7

\bibitem{rasch2009magnetoelastic}
Rasch J~C, Boehm M, Ritter C, Mutka H, Schefer J, Keller L, Abramova G~M,
  Cervellino A and L{\"o}ffler J~F 2009 {\em Phys. Rev. B\/} {\bf 80} 104431

\bibitem{carlsson2011suppression}
Carlsson S, Rousse G, Yamada I, Kuriki H, Takahashi R, L{\'e}vy-Bertrand F,
  Giriat G and Gauzzi A 2011 {\em Phys. Rev. B\/} {\bf 84} 094455

\bibitem{Kobayashi2016}
Kobayashi S, Ueda H, Michioka C and Yoshimura K 2016 {\em Inorg. Chem.\/} {\bf
  55} 7407--7413

\bibitem{sugiyama2018deviation}
Sugiyama J, Nozaki H, Umegaki I, Kobayashi S, Michioka C, Ueda H, Yoshimura K,
  Sassa Y, Forslund O~K, M{\aa}nsson M {\em et~al.\/} 2018 Deviation of
  internal magnetic field in the crse2 triangular lattice with li intercalation
  {\em Proceedings of the 14th International Conference on Muon Spin Rotation,
  Relaxation and Resonance ($\mu$SR2017)\/} p 011004

\bibitem{huang2017layer}
Huang B, Clark G, Navarro-Moratalla E, Klein D~R, Cheng R, Seyler K~L, Zhong D,
  Schmidgall E, McGuire M~A, Cobden D~H {\em et~al.\/} 2017 {\em Nature\/} {\bf
  546} 270--273

\bibitem{bedoya2021intrinsic}
Bedoya-Pinto A, Ji J~R, Pandeya A~K, Gargiani P, Valvidares M, Sessi P, Taylor
  J~M, Radu F, Chang K and Parkin S~S 2021 {\em Science\/} {\bf 374} 616--620

\bibitem{wu2022quasi}
Wu F, Guti{\'e}rrez-Lezama I, L{\'o}pez-Paz S~A, Gibertini M, Watanabe K,
  Taniguchi T, von Rohr F~O, Ubrig N and Morpurgo A~F 2022 {\em Advanced
  Materials\/} {\bf 34} 2109759

\bibitem{Lopez-Paz2022}
López-Paz S~A, Guguchia Z, Pomjakushin V~Y, Witteveen C, Cervellino A,
  Luetkens H, Casati N, Morpurgo A~F and von Rohr F~O 2022 {\em Nature Comm.\/}
  {\bf 13} 4745

\bibitem{Ronneteg2005}
Ronneteg S, Lumey M~W, Dronskowski R and Berger R 2005 {\em J. Alloys Compd.\/}
  {\bf 403} 71--75

\bibitem{Kobayashi2019}
Kobayashi S, Katayama N, Manjo T, Ueda H, Michioka C, Sugiyama J, Sassa Y,
  Forslund O~K, M{\aa}nsson M, Yoshimura K and Sawa H 2019 {\em Inorg. Chem.\/}
  {\bf 58} 14304--14315

\bibitem{song2019soft}
Song X, Cheng G, Weber D, Pielnhofer F, Lei S, Klemenz S, Yeh Y~W, Filsinger
  K~A, Arnold C~B, Yao N {\em et~al.\/} 2019 {\em Journal of the American
  Chemical Society\/} {\bf 141} 15634--15640

\bibitem{gibertini2019magnetic}
Gibertini M, Koperski M, Morpurgo A~F and Novoselov K~S 2019 {\em Nature
  nanotechnology\/} {\bf 14} 408--419

\bibitem{Baenitz2021}
Baenitz M, Piva M~M, Luther S, Sichelschmidt J, Ranjith K~M, Dawczak-Debicki H,
  Ajeesh M~O, Kim S~J, Siemann G, Bigi C, Manuel P, Khalyavin D, Sokolov D~A,
  Mokhtari P, Zhang H, Yasuoka H, King P~D, Vinai G, Polewczyk V, Torelli P,
  Wosnitza J, Burkhardt U, Schmidt B, Rosner H, Wirth S, K{\"{u}}hne H, Nicklas
  M and Schmidt M 2021 {\em Phys. Rev. B\/} {\bf 104} 1--16

\bibitem{Engelsman1973}
Engelsman F, Wiegers G, Jellinek F and {Van Laar} B 1973 {\em J. Solid State
  Chem.\/} {\bf 6} 574--582

\bibitem{Bongers1968}
Bongers P, {Van Bruggen} C, Koopstra J, Omloo W, Wiegers G and Jellinek F 1968
  {\em J. Phys. Chem. Solids\/} {\bf 29} 977--984

\bibitem{VanLaar1973}
{Van Laar} B and Engelsman F~M 1973 {\em J. Solid State Chem.\/} {\bf 6}
  384--386

\bibitem{Nocerino2022a}
Nocerino E, Kobayashi S, Witteveen C, Forslund O, Matsubara N, Tang C,
  Matsukawa T, Hoshikawa A, Koda A, Yoshimura K {\em et~al.\/} 2022 {\em arXiv
  preprint arXiv:2211.06864\/}

\bibitem{Canfield2016}
Canfield P~C, Kong T, Kaluarachchi U~S and Jo N~H 2016 {\em Philos. Mag.\/}
  {\bf 96} 84--92

\bibitem{Stokes}
Stokes H~T, Hatch D~M and Campbell B~J {ISODISTORT, ISOTROPY Software Suite}
  \urlprefix\url{https://iso.byu.edu/}

\bibitem{Campbell2006}
Campbell B~J, Stokes H~T, Tanner D~E and Hatch D~M 2006 {\em J. Appl.
  Crystallogr.\/} {\bf 39} 607--614

\bibitem{Jeschke2015}
Jeschke H~O, Salvat-Pujol F, Gati E, Hoang N~H, Wolf B, Lang M, Schlueter J~A
  and Valent\'{\i} R 2015 {\em Phys. Rev. B\/} {\bf 92}(9) 094417

\bibitem{Jeschke2019}
Jeschke H~O, Nakano H and Sakai T 2019 {\em Phys. Rev. B\/} {\bf 99}(14) 140410

\bibitem{Koepernik1999}
Koepernik K and Eschrig H 1999 {\em Phys. Rev. B\/} {\bf 59}(3) 1743--1757

\bibitem{Perdew1996}
Perdew J~P, Burke K and Ernzerhof M 1996 {\em Phys. Rev. Lett.\/} {\bf 77}(18)
  3865--3868

\bibitem{Liechtenstein1995}
Liechtenstein A~I, Anisimov V~I and Zaanen J 1995 {\em Phys. Rev. B\/} {\bf
  52}(8) R5467--R5470

\bibitem{Mizokawa1996}
Mizokawa T and Fujimori A 1996 {\em Phys. Rev. B\/} {\bf 54}(8) 5368--5380

\bibitem{Sun2020}
Sun X, Li W, Wang X, Sui Q, Zhang T, Wang Z, Liu L, Li D, Feng S, Zhong S, Wang
  H, Bouchiat V, {Nunez Regueiro} M, Rougemaille N, Coraux J, Purbawati A,
  Hadj-Azzem A, Wang Z, Dong B, Wu X, Yang T, Yu G, Wang B, Han Z, Han X and
  Zhang Z 2020 {\em Nano Res.\/} {\bf 13} 3358--3363

\bibitem{Nocerino2022}
Nocerino E, Witteveen C, Kobayashi S, Forslund O~K, Matsubara N, Zubayer A,
  Mazza F, Kawaguchi S, Hoshikawa A, Umegaki I, Sugiyama J, Yoshimura K, Sassa
  Y, von Rohr F~O and M{\aa}nsson M 2022 {\em Sci. Rep.\/} {\bf 12} 21657

\bibitem{Rodriguez-Carvajal1993}
Rodr{\'{i}}guez-Carvajal J 1993 {\em Phys. B Condens. Matter\/} {\bf 192}
  55--69

\bibitem{Ghosh2019}
Ghosh P, Iqbal Y, M\"{u}ller T, Ponnaganti R~T, Thomale R, Narayanan R, Reuther
  J, Gingras M~J~P and Jeschke H~O 2019 {\em npj Quantum Mater.\/} {\bf 4} 63

\end{thebibliography}
\bibliographystyle{iopart-num}

\end{document}